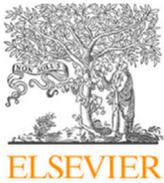
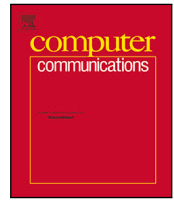

# Learning-based visibility prediction for terahertz communications in 6G networks

Pablo Fondo-Ferreiro [*], Cristina López-Bravo, Francisco Javier González-Castaño, Felipe Gil-Castiñeira, David Candal-Ventureira

*atlanTTic, Universidade de Vigo, Information Technologies Group, Vigo, 36310, Spain*

A B S T R A C T

Terahertz communications are envisioned as a key enabler for 6G networks. The abundant spectrum available in such ultra high frequencies has the potential to increase network capacity to huge data rates. However, they are extremely affected by blockages, to the point of disrupting ongoing communications. In this paper, we elaborate on the relevance of predicting visibility between users and access points (APs) to improve the performance of THz-based networks by minimizing blockages, that is, maximizing network availability, while at the same time keeping a low reconfiguration overhead. We propose a novel approach to address this problem, by combining a neural network (NN) for predicting future user–AP visibility probability, with a probability threshold for AP reselection to avoid unnecessary reconfigurations. Our experimental results demonstrate that current state-of-the-art handover mechanisms based on received signal strength are not adequate for THz communications, since they are ill-suited to handle hard blockages. Our proposed NN-based solution significantly outperforms them, demonstrating the interest of our strategy as a research line.

## 1. Introduction

The growing demands in data traffic require cellular networks to evolve to higher frequencies. Millimeter wave (mmWave) communications have already been considered in 5G networks [1], while terahertz (THz) communications will play a key role for 6G networks, owing to the potential to support terabit-per-second communications [2]. Ultra high THz frequencies, ranging from 0.1 THz to 10 THz, offer tremendous opportunities to attain very high bandwidths in the order of ten to a hundred GHz, solving spectrum scarcity. However, their use for communications introduces new challenges that need to be addressed, including less coverage and extreme sensitivity to blockages even by small-size objects [3].

THz communications are highly vulnerable to blockages, usually implying performance degradation [4]. A channel blockage between the user and its serving access point (AP) renders the communication unfeasible. Regardless of the bandwidth of the channel, the user is no longer able to use it for transmitting while the blockage persists, causing delay and even disconnection events.

Moreover, when a blockage occurs, the user will need to reconnect to a non-blocked AP. The time for this reconfiguration is not negligible and it interrupts the session since the serving AP is not operational. Thus, the relevance of predicting dynamic blockages and self-blockages has been identified as an open challenge to minimize those perceived by the users [4]. Efficient prediction enhances the reliability of the system, required for high-rate low-latency communications [5]. In fact, since attainable bandwidths will even reach the limits of the known approximations of human senses [6], exotic applications such as immersive virtual reality on the move may become feasible. Even though bandwidth may not pose a constraint, reliability issues translate into glitches that can completely destroy user experience.

In order to maintain session continuity in this scenario, some authors have proposed multi-connectivity approaches [7–10] for the users to remain simultaneously connected to multiple APs, thus mitigating the impact of unexpected blockages. However, predicting user blockages would also be relevant even in this case, because blocked channels cannot be used and thus the traffic distribution between the APs must be reconfigured for preventing performance degradation. These redistribution times are less severe if the connections to non-affected APs are kept for as long as possible.

In addition to reconfiguration time, changing the serving AP for a given user also requires exchanging control-plane signaling information (e.g., handover message exchanges, backhaul network reconfiguration commands, etc.), further reducing the efficiency of the network, by introducing overhead. This aspect is relevant to network operators, which will be interested in minimizing the reselections of serving APs for this reason. Summing up, blockages affect network reliability and translate into performance degradation of both the user experience and the system.

In this paper, we study the impact of blockages in ultra-dense THz mobile communication networks, analyze the relevance of accurate predictions of AP visibility for handover decision algorithms to ensure high levels of network availability at all times.

In detail, our main contributions are:

- Characterization of the problem of continuous dynamic association of users to APs in THz communication networks with minimal reconfigurations.
- A mobility management algorithm for making handover decisions based on AP visibility predictions, based on a lightweight NN for predicting future AP visibility from recent past AP visibility measurements, and a visibility probability threshold.
- We evaluate our proposal in terms of network availability and reconfiguration overhead, using different realistic datasets. We compare its performance with state-of-the-art baselines, an optimized 5G handover approach [11], and an idealistic optimal solution based on an oracle. We show that our proposal outperforms them, paving the way for future research on improved visibility prediction algorithms.
- We provide the BloTHz THz network blockage simulator, which can be used to generate realistic visibility datasets in ultra-dense THz networks.

We remark that no previous research has addressed the problem of continuous THz access network availability for mobile users with minimal reconfigurations, as discussed in the next section.

The paper is structured as follows. Section 2 reviews related work. Section 3 presents the problem, Section 4 describes the proposed approach. Section 5 presents and discusses the main results. Finally, Section 6 concludes the paper.

**2. Related work**

Maintaining the reliability and performance of THz communication networks requires overcoming link blockages. Conventional approaches for maintaining session continuity typically involve relocating the user's connection to the AP with the strongest received signal quality [12]. Some degree of hysteresis is applied to reduce unnecessary handovers [13], or additional context variables, such as AP load, are considered to improve network performance [14]. However, these approaches do not address hard blockages as they only consider the current signal strength for handover decisions, rather than the *future* signal strength [15].

Initial solutions to mitigate the effects of link blockages focus on multi-connectivity [16], where terminals keep multiple simultaneous links to multiple APs. In [7] the authors formulated resource allocation optimization problems, focusing on spectrum allocation, and several approaches to solve them. They considered sub-band allocation with varying bandwidths for each user, combined with multi-connections, for improving throughput in comparison with systems without multi-connectivity yet with equal allocation of frequency bands to each user.

In [10], the authors examined the selection of beam pairs for multi-user multi-beam transmissions in dense mmWave networks. They proposed solving the selection problem as a non-cooperative game and concluded that this approach could significantly improve robustness against blockages.

In [17], the authors introduced the Fast Wireless Backhaul (FWB). This architecture allows user equipments (UEs) to maintain simultaneous connections with multiple next-generation node Bs (gNBs). The gNBs are arranged in a multicast tree that receives all the packets destined to a UE. Initially, only one gNB sends packets to the UE. In case of failure, any of the other gNBs can become senders and maintain the data plane connection, as packets are already available in their downlink buffers.

In addition to improving transmission rates or reducing transmission latency, multi-connectivity fosters reliability in critical scenarios, such as emergency situations. In [18], the authors proposed an architecture that integrated the mmWave micro gNBs and sub-6 GHz macro gNBs of a HetNet, allowing UEs (ambulances) to transmit data through both radio access technologies, using dual-connectivity.

Although multi-connectivity is advantageous in terms of reliability as mentioned in Section 1, it also has some drawbacks. Link blockages are not completely avoided, and another issue is excessive resource consumption. For instance, in [17], it was observed that handover overload resulted in decreased available bandwidth in the backhaul.

Some authors have suggested alternative approaches that involve detecting blockages in early stages [19] or even predicting them and then proactively responding [20]. The proactive response may be a beam reconfiguration or a handover decision.

Volatile THz channels, random user mobility patterns, changes in the environment including system hardware (malfunctions, uncalibrated antennas), etc. hinder the beam management process. However, thanks to machine learning algorithms that can learn complex mobility patterns and track environmental dynamics, beams can be managed in complex dynamic scenarios [21]. For instance, in [22] the authors presented a double deep Q-network within a federated learning framework to optimize beam management in an ultradense mmWave scenario.

In [23], a reinforcement learning-based framework was presented for adapting codebook beams based on received power measurements. One of the main advantages of this framework is that it does not require explicit channel knowledge, which can be difficult to obtain in THz systems. It efficiently learns near-optimal beam patterns for highly mobile applications.

Beam tracking and selection alone are not sufficient to solve all cases of link blockage, specially in the case of self-blocking. In such situations, as previously said, handovers may be necessary. Producing accurate blockage predictions to trigger handovers requires input information. This information can be derived solely from in-band data (e.g., the sequence of beams used to serve a mobile user [24]), it can be obtained out-of-band (e.g., information gathered by RGB and RGB-D cameras [25]), or it can be a combination of in-band and out-of-band data, as in [26] (LiDAR and wireless mmWave/sub-THz signatures), or [27] (RGB cameras and wireless mmWave/sub-THz signatures). Obviously, the complexity and the cost of the solution varies with the number of sources and the requirement of dealing with out-of-band information.

In [28] the authors jointly considered handover and beam management processes. They presented a learning solution for THz drones that utilizes a recurrent neural network to predict the serving base station and the serving beam for each drone. These predictions are based on prior observations of drone location and beam trajectories.

Therefore, to mitigate the impact of blockages, most previous works have focused on multi-connectivity approaches to maintain connections with multiple access points. However, these approaches do not eliminate blockages but merely reduce their impact by increasing the cost and complexity of the access points and terminals, and increasing network resource usage. Additionally, existing approaches attempt to determine or predict whether a link between a user and an AP is blocked or not. To the best of our knowledge, no previous work has estimated link visibility *probabilities*, instead of taking hard decisions, for maximizing availability level while avoiding unnecessary reconfigurations. This latter aspect not only lowers energy consumption (which is especially relevant at the terminal side) but also improves network stability, thereby reducing traffic delay and jitter.

In this work we seek to introduce a novel approach to the problem in that line. Instead of predicting link blockages, we propose to predict the probability that an AP will remain visible. This provides a range of visibility probabilities for the set of APs (as opposed to the binary blocked/not-blocked classification in previous works) and allows us to control the number of reconfigurations of network links to keep users connected to the same APs as long as possible. This reduces delay and network overhead, while avoiding excessive use of network resources, thus improving network efficiency.

## 3. Problem statement

We consider a next-generation mobile network composed by $M$ APs[1] providing wireless connectivity to $N$ moving users. This wireless network operates on the THz frequency band, to provide the high data rates demanded by the user applications.

We also consider a time-slotted system model. At each time slot $t$, every user moving across the scenario can be connected to a single AP if within its coverage (that is, the terminals employ single-beam technology). Communication then takes place as long as the user stays in the coverage area of the AP and the corresponding channel is not blocked. The location of user $i$ at time slot $t$ is represented by its Cartesian coordinates, $x_i(t)$ and $y_i(t)$, and its rotation (orientation) relative to axis $x$ is given by angle $\phi_i(t)$. The current location of each user can be directly obtained by the network using existing procedures defined in the 3GPP standard [29], while the orientation of each user can be obtained from inertial sensors in the terminal, or estimated by the network based on signal measurements [30]. Note that network-based estimations may be very accurate in ultra-dense scenarios, given the high density of potentially visible APs in the vicinity of the coverage of each user.

The coverage area of an AP is given by the locations for which its signal strength exceeds a power threshold. The path loss at distance $d$ from the AP is calculated with the log-distance path loss model for indoor THz scenarios in [31], as given by Eq. (1).

$$\overline{PL}(d) = PL(d_0) + 10\gamma \log\left(\frac{d}{d_0}\right) + X_\sigma. \tag{1}$$

$\overline{PL}(d)$ is the average path loss (in dB) at distance $d$, $PL(d_0)$ is the path loss at reference distance $d_0$, $\gamma$ is the path loss exponent and $X_\sigma$ is a Gaussian random variable with zero mean and standard deviation $\sigma$ (in dB). We will consider that a terminal lies inside the coverage range of an AP if the pathloss for the distance in between (given by Eq. (1)) is lower than threshold $P_{th}$.

Due to the characteristics of THz communication beams, only line-of-sight (LOS) communications are feasible, as non-line-of-sight (NLOS) signals experience strong attenuation [3]. Thus, a blockage happens when there is no direct visibility between the terminal and the AP. We define the binary input matrix $\mathbf{V(t)} = \{v_{ij}(t)\}$, where $v_{ij}(t) \in \{0, 1\}$ is the visibility between terminal $i$ and AP $j$ at time slot $t$. Note that $v_{ij}(t) = 1$ if, at time slot $t$, the channel between user $i$ and AP $j$ is not blocked *and* user $i$ lies inside the coverage of AP $j$. No data can be transmitted at time slot $t$ between user $i$ and AP $j$ when $v_{ij}(t) = 0$. Two types of blockages can happen in our scenario: self-blockages, in which the body of the person carrying the mobile terminal causes the blockage; and external blockages, caused by the bodies of other people around. This can be the case of an open indoor space or an urban pedestrian street, where deploying highly dense THz APs may be cost-effective [32]. Even though we are using an indoor scenario as an example in our problem statement, the proposed solution is directly applicable to outdoor environments without additional considerations.

In this scenario, our goal is to dynamically determine the assignments of APs to users for the next time slot ($t + 1$), maximizing network availability by avoiding blockages. To this end, we define binary matrix $\mathbf{A(t)} = \{a_{ij}(t)\}$, where $a_{ij}(t) \in \{0, 1\}$ represents whether user $i$ is associated to AP $j$ at time slot $t$. Overall, the objective function component to be maximized for the next time slot $t + 1$ is given by Eq. (2).

---

[1] We also refer to the APs as base stations (BSs) throughout the manuscript.

**Table 1**
Input data of the problem.

| | |
|---|---|
| $N$ | Number of users (i.e., mobile terminals). |
| $M$ | Number of APs. |
| $x_i(t)$ | $x$-axis Cartesian coordinate of user $i$ location at time slot $t$. |
| $y_i(t)$ | $y$-axis Cartesian coordinate of user $i$ location at time slot $t$. |
| $\phi_i(t)$ | Rotation of user $i$ around axis $z$ at time slot $t$. |
| $v_{ij}(t)$ | Binary variables that equal 1 if user $i$ and AP $j$ are mutually visible at time slot $t$ (i.e., the user lies within AP coverage and the channel is not blocked). |
| $a_{ij}(t)$ | Binary variables that equal 1 if user $i$ is connected to AP $j$ at time slot $t$. |

**Table 2**
Output decision variables of the problem.

| | |
|---|---|
| $a_{ij}(t+1)$ | Binary variables that equal 1 if user $i$ is scheduled to be connected to AP $j$ at time slot $t + 1$. |

$$f_1(\mathbf{A}(t+1), \mathbf{V}(t+1)) = \sum_{i=1}^{N} \sum_{j=1}^{M} a_{ij}(t+1) \cdot v_{ij}(t+1) \tag{2}$$

We normalize network availability by dividing the value provided by Eq. (2) by the number of users $N$, as given by Eq. (3).

$$\tilde{f}_1(\mathbf{A}(t+1), \mathbf{V}(t+1)) = \frac{f_1(\mathbf{A}(t+1), \mathbf{V}(t+1))}{N} \tag{3}$$

Besides the constraint that $a_{ij} \in \{0, 1\}$ is a binary variable, we must take into account the constraint that, at any time slot, each user can only be assigned a single AP (4).

$$\sum_{j=1}^{M} a_{ij}(t) = 1, \ \forall i \in \{1, \ldots, N\} \tag{4}$$

Note that the current association between users and APs (i.e., $a_{ij}(t)$ $\forall i, j$) is a set of input variables of the problem, while the association for the next slot (i.e., $a_{ij}(t+1)$ $\forall i, j$) is the set of output decision variables. Moreover, the visibility at time slot $t + 1$ (i.e., $\mathbf{V}(t+1)$) is unknown at time slot $t$.

In addition, we are also interested in performing as few reassignments (i.e., changes in serving APs) as possible due to the energy consumption, overhead and instability introduced by reconfigurations. A reconfiguration for user $i$ at time slot $t$ is denoted by binary variable $r_i(t) \in \{0, 1\}$ (5).

$$r_i(\mathbf{A}(t), \mathbf{A}(t-1)) = \frac{1}{2} \sum_{j=1}^{M} \left| a_{ij}(t) - a_{ij}(t-1) \right| \tag{5}$$

Thus, the reconfiguration component of the objective function to be minimized for the next time slot $t + 1$ is given by Eq. (6).

$$f_2(\mathbf{A}(t+1), \mathbf{A}(t)) = \sum_{i=1}^{N} r_i(\mathbf{A}(t+1), \mathbf{A}(t)) \tag{6}$$

We also normalize this value by the number of users $N$, yielding (7).

$$\tilde{f}_2(\mathbf{A}(t+1), \mathbf{A}(t)) = \frac{f_2(\mathbf{A}(t+1), \mathbf{A}(t))}{N} \tag{7}$$

Tables 1 and 2 summarize the problem's input data and output decision variables of the considered system model, respectively.

## 4. Proposed solution

We propose a solution to the user-to-AP association decision problem based on a prediction of future APs' visibilities and a selection criterion. This solution is not tailored to a specific architecture but different deployment scenarios can be considered, including both on the user side (i.e., UE) and on the network side (e.g., APs). Interestingly, as we will show in Section 5, the information provided by considering user positions does not improve the overall performance of the model, so our final model only considers AP visibilities as input parameters, which simplifies the implementation.

The different deployment options introduce some trade-offs that must be considered, in terms of signaling overhead, added delay in the AP selection process and energy consumption of end devices. If the decision process is executed on end devices, it does not involve any signaling overhead since AP visibilities are known to the UEs. However, the increase in energy consumption of mobile terminals must be considered and the limited computing capabilities may introduce a delay in the AP selection process. If the decision process is executed on the network side (e.g., AP), there will be some signaling overhead (e.g., visibility measurements reported from the UEs to the APs). However, the delay in AP selection will be reduced thanks to the increased computing capabilities at the network side. Moreover, energy usage at end devices will not increase. Note that this last deployment option follows the same architecture defined by the 3GPP for handover in cellular networks.

The proposed approach consists in two sequential steps:

1. Prediction of future APs' visibilities.
2. AP selection.

In the first step, we predict the future APs' visibilities for the next time slot, based on the information that is available at the current time slot. In the second step, based on the predicted visibility information, we propose a heuristic algorithm for selecting the AP each user should be granted at the next time slot. Each step is described in detail in the following subsections.

### 4.1. Prediction of future APs' visibilities

We address the estimation of the probabilities of future APs' visibilities with a feed-forward NN. This NN predicts, for a given user terminal, the probability that each AP will be visible at the following time-slot.

The input variables of the NN for user $i$ at time slot $t$ include the following information from the previous $H$ time slots and the current one:

1. The visibilities between user $i$ and each AP $j$, given by

   $v_{ij}(t-\delta) \; \forall \; j \in [1,\ldots,M],$
   $\delta \in [1,\ldots,H].$

2. The location of user $i$, given by

   $x_i(t-\delta), y_i(t-\delta) \; \forall \; \delta \in [1,\ldots,H].$

3. The rotation of user $i$, given by

   $\phi_i(t-\delta) \; \forall \; \delta \in [1,\ldots,H].$

4. The location of every other user $k \neq i$, given by

   $x_k(t-\delta), y_k(t-\delta) \; \forall \; k \in [1,\ldots,N] \; | \; k \neq i,$
   $\delta \in [1,\ldots,H].$

5. The rotation of every other user $k \neq i$, given by

   $\phi_k(t-\delta) \; \forall \; k \in [1,\ldots,N] \; | \; k \neq i,$
   $\delta \in [1,\ldots,H].$

In Section 5 we explore the impact of each of these input variables in the visibility prediction performance, in order to select the simplest and most generalizable model.

The target output variables that the NN predicts are the future visibility probabilities of the APs from the users $p_{ij}(t+1) \; \forall \; i \in [1,\ldots,N] \; \forall \; j \in [1,\ldots,M]$ at time $t+1$ given by

$$p_{ij}(t+1) = \Pr\left(v_{ij}(t+1) = 1\right). \tag{8}$$

Note that despite the model intrinsically supports sudden blockages and mobility patterns, the NN model needs to be retrained if there are relevant changes in the underlying environment (e.g., changes in the base stations, new mobility patterns, …).

### 4.2. AP selection

The objective of the AP selection step is to determine which AP should be selected at the next time slot for every user, based on the outcome of the visibility prediction step, that is, the predicted AP visibility probabilities at the next time slot.

The simplest alternative would be to select for every user the AP with the highest predicted probability to be visible. However, although this provides the highest availability for a given prediction, it can introduce a significant increase of reconfiguration overhead as given by Eq. (7). Hence, in order to reduce it, we apply a hysteresis mechanism for selecting the *most-likely-to-be-available* AP or keeping the currently assigned one. Specifically, we propose to keep the latter if the difference between the highest probability and the probability for the current AP to stay visible is less than a given threshold $T$; otherwise, the AP with the highest visibility probability is selected. For the sake of completeness, at the first time slot, the AP with the highest predicted visibility probability is selected. For subsequent time slots, the output matrix $\mathbf{A}(t+1)$ is given by:

$$a_{ij}(t+1) = \begin{cases} a_{ij}(t) & \text{if } p_{ij^*}(t+1) - p_{ij}(t+1) < T \\ \begin{cases} 1 & \text{if } j = j^* \\ 0 & \text{if } j \neq j^* \end{cases} & \text{otherwise} \end{cases}$$

where $j^*$, given by Eq. (9), is the index of the AP with highest predicted visibility probability for user $i$ at time slot $t+1$.

$$j^* = \arg\max_j \Pr\left(v_{ij}(t+1) = 1\right). \tag{9}$$

## 5. Results

### 5.1. Blockage simulator for dataset generation

There exist simulators for the physical layer of THz communications such as TeraSim [33], but they are not suitable for our purposes because they do not consider blockages caused by obstacles. Hence, in this work we have developed a novel Unity-based 3D simulator, BloTHz, for generating AP visibility and blockage datasets for THz wireless networks in scenarios with user mobility (BloTHz blockage simulator in the sequel). We have made it publicly available at [34]. The BloTHz blockage simulator considers a 3D rectangular prismatic room of configurable dimensions with multiple APs providing connectivity to an also configurable number of users. By default, the APs are arranged at the crossings of a square grid mesh on the ceiling of the room, with configurable cell edge size (this layout has been considered to minimize blockages in previous research on indoor THz coverage [7,35], and in standards [36]). The BloTHz blockage simulator also allows individual APs to be deployed in specific locations. The users moving in the scenario are modeled as upright objects on the floor with human body shapes, with two heights, 1.60 and 1.80 m. Each user has a mobile terminal placed 20 cm in front of his/her chest. Throughout the simulation, mobile terminals keep that relative position while moving with the persons carrying them. For illustrative purposes, the BloTHz blockage simulator also provides a graphical user interface (GUI) for visually inspecting simulation states. Fig. 1 shows two screenshots of the simulator showing a direct visibility and a blockage scenario, respectively.

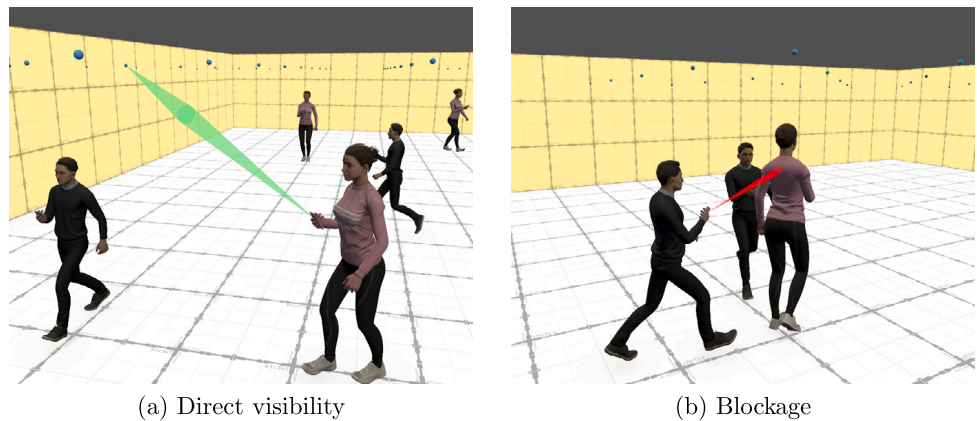

**Fig. 1.** Screenshot of the BloTHz simulator for dataset generation.

Blockages are determined by considering a configurable beam width that represents the aperture angle of the transmitter and receiver antennas. We consider that the beams propagate through the 3D volume resulting from the intersection of two cones, respectively originated at the transmitter and receiver antennas, whose base radii grow with distance depending on the given aperture angles. The axes of the cones are perfectly aligned, by considering that the beams of the transmitter and receiver antennas are configured to point to each other. Blockages are caused by user bodies, including self-blockages and those caused by other users. The blockage simulator considers that a blockage takes place when the region given by the union of the cones by their bases is completely intercepted by a user (see Fig. 1.b). This is approximated with a ray tracing algorithm with a finite number of rays inside the region, including the straight line containing the axes of the cones, which connects both transmission ends.

Realistic user mobility has been implemented as follows: Users are initially deployed at random floor positions and then start moving in a straight line at constant speed towards respective random generated positions inside the room. Once each user reaches the target position, he/she waits a uniformly distributed random time in [0.5, 5] s. Then, a new target destination is set at random, the body of the user rotates in $z$ to face it, and the user resumes straight movement in that direction. This process is repeated until the end of the simulation. During the movement of the user, the mobile phone is always headed towards the target location, representing a typical situation in which the user manipulates the terminal while walking.

The states of the simulation are logged at configurable intervals (time slots) as a time series of comma-separated values (CSV) in a text file. This includes the 3D positions of every user, his/her rotations around the $z$ axis, and the distances between the terminal and each AP if the corresponding link is not blocked (or a blocking flag otherwise). Note that a non-blocked channel between a user and an AP does not necessarily imply that the user lies within coverage of the AP, as the received signal strength fades with distance. Saved distances between the user and the non-blocked APs allow considering channel loss models for different wireless technologies. In this paper, we determine coverage ranges from distance measurements based on Eq. (1), and coverage limit is set to $P_{th}$, as previously said.

### 5.2. Approaches in the evaluation

In order to evaluate the outcome of our proposal, we compared it with state-of-the-art signal hysteresis-based handover, a naïve visibility baseline, an optimized 5G handover proposal from the literature [11], and an ideal optimum prediction.

Let us remind that our proposal, described in Section 4, consists in first predicting the AP visibility probability using a NN. Then, the *most-likely-to-be-available* APs are selected if the corresponding probabilities are higher than threshold $T$, compared to the visibility probabilities of the currently assigned APs. Otherwise, the current ones are kept.

Signal hysteresis-based handover is the typical mechanism used in cellular networks, where UEs take handover decisions based on the quality of the received signal [37] if a hysteresis threshold is exceeded. Selected APs will be those with the highest received signal power values, if those values minus the threshold are stronger than the signal of the currently assigned APs.

The naïve visibility baseline just considers that the visibility for the next time slot will be identical to that in the current time slot. Based on that naïve prediction, the baseline selects the currently assigned APs at the next time slot if still available. Otherwise, it randomly selects APs for disconnected terminals between all available APs.

The optimized 5G handover approach implements the Individualistic Dynamic Handover Parameter Optimization algorithm based on an Automatic Weight Function (IDHPO-AWF) proposed in [11]. The handover decision is based on the AP with the highest received signal strength, which must exceed that of the serving AP by a handover margin (HOM) for a consecutive time-to-trigger (TTT) period. In the IDHPO-AWF approach, the HOM and TTT values are dynamically estimated for each UE.

The ideal optimum prediction approach assumes that the visibility for the next time slot is known (i.e., provided by an oracle). Based on that ideal knowledge, it keeps the currently assigned APs at the next time slot if still available. Otherwise, it randomly selects APs for disconnected terminals between all available APs, as in the previous case.

### 5.3. Metrics

We compared our proposal with the alternatives using the following performance metrics:

- Normalized network availability $\tilde{f}_1$ given by Eq. (3), the fraction of users whose assigned APs at a time slot are visible and available (i.e., within coverage and not blocked).
- Normalized reconfiguration overhead $\tilde{f}_2$ given by Eq. (7), the fraction of users whose assigned AP changes with respect to the previous time slot.

### 5.4. Pathloss parameters and NN hyperparameters

The values for the different parameters in Eq. (1) were taken from [31] for frequencies between 0.14 and 0.22 THz. In our experiments, we chose a central frequency of 0.14 THz, $\gamma = 2.117$, $\sigma = 0.5712$, $d_0 = 0.35$ m and $PL(d_0) = 25$ dB. The pathloss threshold for AP coverage was set to $P_{th} = 55$ dB, corresponding to around 8 m.

The hyperparameters of the NN were experimentally adjusted to attain a good prediction accuracy without overfitting. Table 3 presents the final selection.

**Table 3**
Hyperparameters of the neural network.

| | |
|---|---|
| Number of hidden layers | 1 |
| Number of neurons in hidden layer | 1000 |
| Activation function for the hidden layer | Rectified linear unit (ReLU) |
| Number of training epochs | 750 |
| Training algorithm | Stochastic Gradient Descent (SGD) |
| Loss function | Binary Cross Entropy |

*5.5. Experiments*

We conducted different experiments to evaluate the performance of the proposed solution departing from different datasets generated with the BloTHz blockage simulator described in Section 5.1. BloTHz settings were a $400 \text{ m}^2$ square room with a side of $20 \text{ m}$ and a height of $2.4 \text{ m}$, in which the APs were deployed on the ceiling forming a grid with $2 \text{ m}$ cell edge, yielding a total of 121 APs. The aperture angle of the antennas was set to 2.5° [38] and the time slot interval to $0.2 \text{ s}$. The number of users in the scenario varied from 1 to 10 in the different experiments. We considered different simulation durations to generate the datasets, 100, 1000 and $10\,000 \text{ s}$. Every dataset was split as follows: the first 70% of the observations for training, the next 10% for validation and the final 20% as a testing set. The metrics were calculated for each time slot of this testing set and were averaged as the final results provided with 95th-percentile confidence intervals.

The generated datasets were processed with a second transmission simulator written in Python, which we have released under an open-source license in [39] (THz visibility prediction and AP assignment simulator in the sequel). This simulator takes the slot-by-slot visibility and mobility information in the input dataset, transforms the visibility and distance measurements into received signal strength values according to Eq. (1), implements the approaches described in Section 5.2, and calculates the metrics in Section 5.3, which we used to compare the different approaches.

We performed the following experiments:

1. Relevance of input variables for visibility prediction for different dataset durations, using our NN-based approach.
2. Relevance of past history $H$ for visibility prediction for different dataset durations, using our NN-based approach.
3. Impact of the threshold value ($T$) in the AP selection algorithm, using our complete proposal.
4. Performance of our proposal in comparison with the alternative approaches as the number of users increases.
5. Impact of the number of users in the training set for different number of users in the test set.

The first three experiments are considered tuning experiments with partial settings, the fourth experiment provided the final results and the fifth experiment analyzes the generalizability of the proposed solution. The goal of the first experiment was to obtain insights on the relevance of the input variables of the NN for the visibility prediction sub-problem, using the information from the previous time slot (i.e., $H = 1$), for different dataset durations, and for only five users in the scenario. Table 4 shows its results.

In this first experiment we explored different subsets of input variables for the AP visibility prediction problem, taken from those described in Section 4.1. The first column of the table shows them: AP visibility for user $i$ ($v_{ij}(t-1) \forall j \in [1, \ldots, M]$) (subset #1 in the sequel), location of user $i$ ($x_i(t-1), y_i(t-1)$), rotation of user $i$ ($\phi_i(t-1)$), location of every other user ($x_k(t-\delta), y_k(t-1) \forall k \in [1, \ldots, N] \mid k \neq i$), and rotation of every other user ($\phi_k(t-1) \forall k \in [1, \ldots, N] \mid k \neq i$). First we can observe that the precision, recall and network availability improved as the duration (thus the size) of the datasets increased. Despite for low dataset durations, such as $100 \text{ s}$, precision and recall values are lower than 83%, and the impact on network availability is moderate, as it keeps above 94%. Moderate dataset durations, such as $1000 \text{ s}$, raise the precision and recall to 92%, while network availability reaches almost the maximum value above 98.1%. Further increasing the dataset duration to $10\,000 \text{ s}$ pushes the precision and recall close to 94%, but the improvement in network availability is negligible, reaching a value of 98.3%. This shows that the proposal is robust even for moderate values of precision and recall, so prediction quality has limited impact in performance.

For a given dataset duration, the results were very similar regardless of the subset of variables considered, but the differences were higher for the shortest datasets, as it could be expected. Once the duration of the training dataset was long enough, the performance metrics tended in all cases to the same values.

Interestingly, in this first tuning experiment, the insight was that adding additional input variables to the NN beyond the AP visibilities from the previous time slot did not seem to improve the quality of the predicted values. We can speculate that user positions and rotations are strongly related to AP visibility, making these somewhat redundant. The simplification of the model owing to this decision is beneficial, as we will later discuss. Summing up, for the rest of the experiments in this section, we used AP visibilities as the only input variable for the NN.

It is important to observe that, in addition to the improvements in precision and recall with the increased duration of the datasets, the improvement in availability (which should approach 100%) from 94% to 98% with a predictive algorithm was statistically significant and encouraging.

In the second tuning experiment we studied the relevance of history parameter $H$ for visibility prediction, also for five users and different dataset durations, only for variable subset #1 based on the outcome of the previous experiment. Table 5 shows the results.

The impact in network availability of extending $H$ beyond the previous slot was negligible. The same levels of precision, recall and network availability were obtained up to the last five slots, for a slight improvement in precision and recall (less than 1% in both cases).

Given the results of the first two experiments, in the third tuning experiment the duration of the datasets was set to $10\,000 \text{ s}$, $H$ was set to 1 and variable subset #1 was used. This experiment analyzed the impact of threshold $T$ in the performance of the AP selection algorithm. Fig. 2 shows the network availability and reconfiguration overhead for different threshold values for the scenario with five users.

Network availability gradually decreased as the probability threshold $T$ increased, from over 99% for $T < 0.05$ to 90% for $T > 0.9$. For $T < 0.6$, the decrease seemed nearly linear, down to an availability of ~0.96 at that point. The decrease rate was then intensified for $0.6 \leq T < 0.9$. The impact on reconfiguration overhead was significantly different. Reconfiguration was considerably high for very low probability threshold values (e.g., $T < 0.01$) but abruptly dropped as the threshold increased. The normalized reconfiguration overhead exceeded 0.3 for $T = 0$, but it was lower than 0.1 for $T = 0.01$. Then, it kept smoothly decreasing to reach ~0.05 for $T = 1$. According to these results, the probability threshold allows modulating the performance of our solution by regulating the trade-off between network availability and reconfiguration overhead, as desired. The exact threshold value might require some fine tuning for different scenarios. Very low values result in too frequent reconfigurations, introducing a substantial overhead. On the other hand, high threshold values result in lower responsiveness to blockages, thus reducing network availability. Seeking a good performance in both metrics, low values of $T$ in the $[0.01, 0.1]$ range should be selected. In the sequel we considered $T = 0.05$ a good compromise.

Building on these insights, the fourth experiment explored the performance of the proposed approach in comparison with other strategies as the number of users increased. Fig. 3 shows the respective values for network availability and reconfiguration overhead for an increasing number of users.

Table 4
Relevance of input variables of the NN for visibility prediction (first experiment).

| Variables | Dataset duration (s) | Precision | Recall | Network availability |
|---|---|---|---|---|
| $v_{ij}(t-1) \ \forall \ j \ \in \ [1,\ldots,M]$ | 100 | 0.819 | 0.820 | 0.941 |
|  | 1000 | 0.921 | 0.918 | 0.981 |
|  | 10000 | **0.937** | **0.939** | **0.983** |
| $v_{ij}(t-1), x_i(t-1), y_i(t-1) \ \forall \ j \ \in \ [1,\ldots,M]$ | 100 | 0.822 | 0.827 | 0.963 |
|  | 1000 | 0.921 | 0.919 | 0.982 |
|  | 10000 | 0.937 | 0.939 | 0.983 |
| $v_{ij}(t-1), x_i(t-1), y_i(t-1), \phi_i(t-1), \ \forall \ j \ \in \ [1,\ldots,M]$ | 100 | 0.825 | 0.827 | 0.963 |
|  | 1000 | 0.921 | 0.919 | 0.981 |
|  | 10000 | 0.937 | 0.939 | 0.983 |
| $v_{ij}(t-1) \ \forall \ j \ \in \ [1,\ldots,M]$ $x_k(t-1), y_k(t-1), \phi_k(t-1) \ \forall \ k \in [1,\ldots,N]$ | 100 | 0.830 | 0.815 | 0.965 |
|  | 1000 | 0.920 | 0.921 | 0.981 |
|  | 10000 | 0.937 | 0.939 | 0.983 |

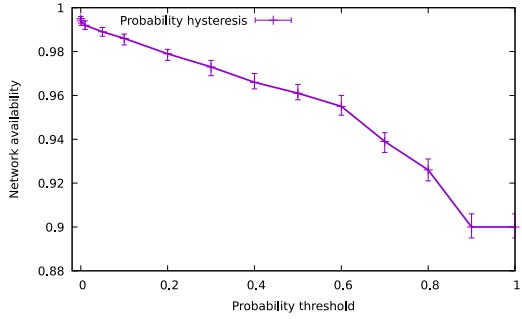

(a) Network availability

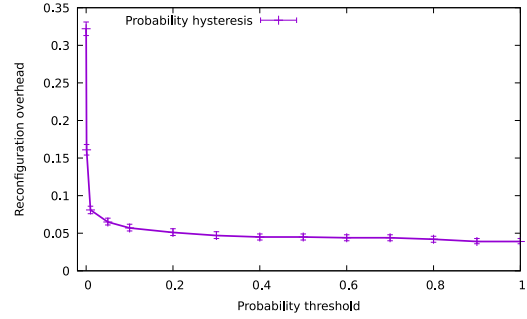

(b) Reconfiguration overhead

**Fig. 2.** Impact of threshold $T$ of the AP selection algorithm of our proposal in terms of network availability and reconfiguration overhead.

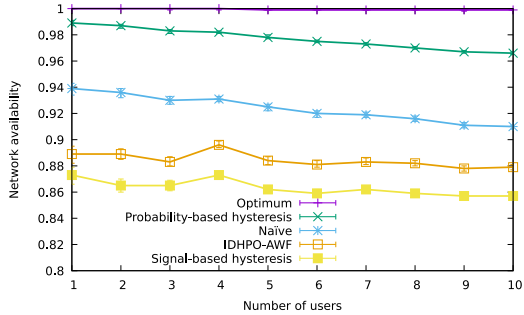

(a) Network availability

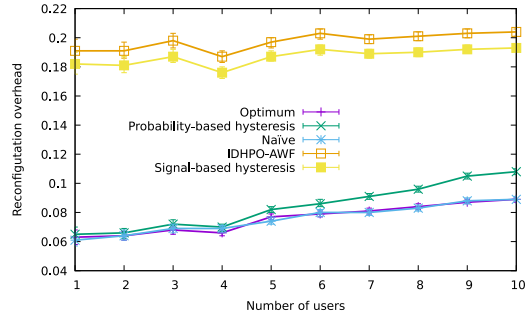

(b) Reconfiguration overhead

**Fig. 3.** Network availability and reconfiguration overhead for different approaches.

The figure shows a slight decrease in network availability as the number of users grew. This was true for every approach but for the optimal oracle-based approach, for which a 100% network availability was always feasible. This is relevant because it means that 100% availability would be achievable in the scenarios if an availability oracle were available. Our proposed probability-based approach with hysteresis performed much closer to the optimal than the remaining alternatives, achieving over 98% network availability for less than 5 users. Even for 10 users, network availability remained above 96%. The best results attained by the naïve approach were less than 94% for a single user, dropping to 91% for 10 users, which is clearly unacceptable for many use cases. Finally, the IDHPO-AWF optimized 5G handover [11] and the baseline signal hysteresis-based approach obtained the worst results, less than 90% and 88% network availability, respectively. These results drop to 88% and 86% for 10 users, respectively, that is, over 10% penalty compared to our proposed probability-based approach with hysteresis based on NN predictions.

Reconfiguration overhead also grew with the number of users. The optimum and naïve approaches behaved nearly identically, obtaining the lowest values for this metric as expected, since both approaches kept previous AP assignments for as long as they remained visible. The values varied between 0.06 for one user and 0.09 for ten users. Our approach followed a similar pattern, yet with slightly increased reconfiguration overheads (about 0.11 for ten users). As the signal hysteresis-based approach and the IDHPO-AWF proposal ignored reconfiguration overhead, their resulted in substantially more frequent AP re-assignments. Specifically, reconfiguration overhead ranged between 0.18 and 0.2 in the former and between a 0.19 and 0.21 in the latter, doubling that of the other approaches.

Finally, the fifth and last experiment explored the impact on the performance of the number of users in the training set for different numbers of users in the test set. Fig. 4 shows the network availability of the proposed solution using three models trained with 1, 5 and 10 users, respectively, and evaluated with different test sets ranging from 1 to 10 users.

We can observe that the three models perform very similarly, with a difference below 0.5%. The model trained with 10 users achieves the best network availability for every number of users in the test set, followed by the model trained with 5 users and lastly by the model with 1 user. This result is indicative of the generalizability of our proposed solution and its adaptability to changing conditions in the number of users in the environment.

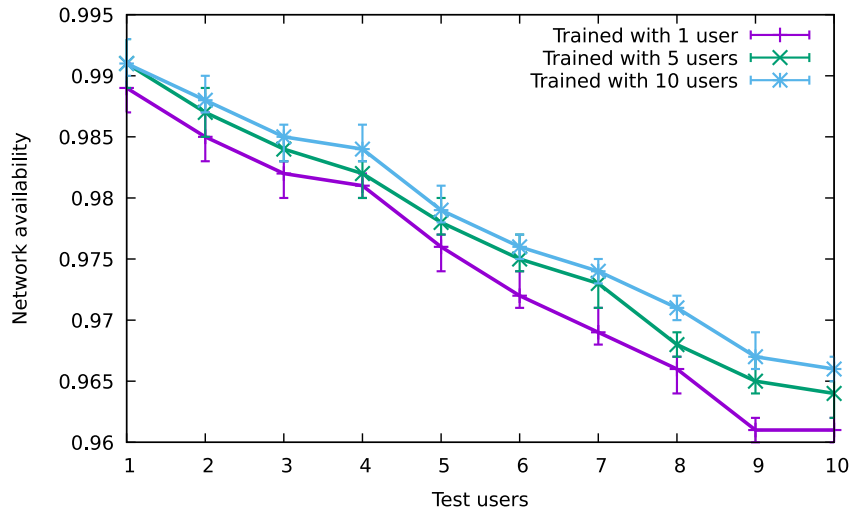

**Fig. 4.** Network availability for different numbers of training users and test users.

**Table 5**
Relevance of the $H$ parameter for visibility prediction with the NN (second experiment).

| $H$ | Dataset duration (s) | Precision | Recall | Network availability |
|---|---|---|---|---|
| 1 | 100 | 0.819 | 0.820 | 0.941 |
|   | 1000 | 0.921 | 0.918 | 0.981 |
|   | 10000 | 0.937 | 0.939 | **0.983** |
| 2 | 100 | 0.841 | 0.844 | 0.967 |
|   | 1000 | 0.921 | 0.919 | 0.980 |
|   | 10000 | 0.937 | 0.944 | 0.983 |
| 3 | 100 | 0.850 | 0.857 | 0.939 |
|   | 1000 | 0.920 | 0.921 | 0.979 |
|   | 10000 | 0.937 | **0.946** | 0.983 |
| 4 | 100 | 0.844 | 0.853 | 0.947 |
|   | 1000 | 0.919 | 0.921 | 0.982 |
|   | 10000 | 0.938 | 0.946 | 0.982 |
| 5 | 100 | 0.843 | 0.852 | 0.947 |
|   | 1000 | 0.920 | 0.921 | 0.980 |
|   | 10000 | **0.939** | 0.946 | 0.982 |

*5.6. Discussion*

Our two-stage simulation results (including the BloTHz simulator for blockage dataset production and a THz visibility prediction and AP assignment simulator) show that the signal hysteresis-based approach may weaken the performance in THz networks due to hard blockages. In the scenarios considered, if an oracle were available to accurately predict blockages, it would be possible to achieve almost a 100% network availability, but the signal-hysteresis based approach cannot even reach 90% availability. The main reason for this performance drop may be due to the fact that the APs with the strongest received signal may be too close to the user. Since users are continuously moving, it is likely that a user will move past an AP and incur in self-blockage, triggering continuous connection reconfigurations to alternative non-blocked APs. Consequently, the signal hysteresis-based approach has a ~12% increase in reconfiguration overhead compared to the optimum approach.

Interestingly, a naïve assignment by maintaining the current user–AP connections for as long as possible, followed by random assignment to visible APs when necessary, consistently attains better performance. In this work, seeking a trade-off between visibility and reconfiguration overhead, we propose a more elaborated approach, consisting of two sequential stages, by first predicting the user-to-AP visibility probability using an NN and then selecting the most likely candidate to be available if a probability threshold is exceeded. This approach shows promise to achieve nearly optimal results, demonstrating the value of predictive algorithms to reach the trade-off. The approach outperforms the naïve strategy by 5% and the baseline signal hysteresis-based approach by over 10% in terms of network availability. Furthermore, its probability-based hysteresis mechanism keeps reconfiguration overhead very low, close to the optimum.

The initial tuning experiments revealed the low relevance of some input variables for the NN (e.g., user location, rotation). The most meaningful variable for predicting future AP visibility was recent AP visibility, and considering a long history of time slots did not improve the overall quality of the prediction. These insights simplify the implementation of our proposal, by relieving the network from obtaining user positions and rotation from terminal-side sensors, for instance. Therefore, our final proposed solution is simplified to only consider recent AP visibilities as the only input variables.

Moreover, the results of the last experiment suggest the robustness and scalability of our proposal in terms of number of UEs, showing that a model trained with a single user can still provide good performance in an environment with 10 users. On the other hand, the number of APs directly influences the dimension of the input and output layers of the NN, since these represent AP visibilities for a given user. Consequently, the complexity of the model will be directly related to the number of APs. However, this will not have a significant impact in the real-time operation of the proposed solution, since the online prediction phase will still be executed nearly instantaneously. The impact of increasing the number of APs will be noticeable in the offline training phase, which will demand more computing resources and require additional training time. To overcome this potential scalability limitation, paradigms such as Multi-Access Edge Computing (MEC) and cloud computing can be leveraged to offload model training.

Finally, it is worth noting that the probability threshold of the AP selection algorithm can be used to fine tune the trade-off between network availability and reconfiguration overhead. In our experiments, a value of $T = 0.05$ achieved a good trade-off, providing nearly optimal network availability with significantly low reconfiguration overhead.

**6. Conclusions**

THz communications are a key enabler to solve spectrum scarcity and enable new applications in next-generation 6G networks. However, they are highly sensitive to link blockages, which can disrupt ongoing communications. Therefore, continuous network availability, rather than spectrum sharing optimization, becomes a primary goal. In this work we have proposed a visibility prediction solution to assign APs with minimal blockages while keeping a low reconfiguration overhead, thus achieving a good system performance trade-off. Our proposal employs an NN to predict AP visibility probabilities for all APs. The serving AP for a terminal is reconfigured to the *most-likely-to-be-available* AP if its visibility probability exceeds that of the current AP by a threshold.

We have evaluated our proposal on realistic visibility datasets generated with a THz blockage simulator, BloTHz, which we make publicly available in this paper. Our experimental results reveal that existing signal hysteresis-based approaches for handover management perform poorly in THz scenarios due to hard blockages. Conversely, our predictive solution can maintain network performance close to optimal levels, without introducing unnecessary reconfigurations. Moreover, the most relevant parameter for predicting future visibility seems to be the current AP visibility of each user, while positional parameters such as terminal-side user location estimations and rotations do not improve prediction results.

Summing up, we have shown the potential of visibility prediction as a promising approach for exploiting the full potential of THz communication systems by minimizing blockages, thus opening an interesting research line for future work.

**CRediT authorship contribution statement**

**Pablo Fondo-Ferreiro:** Writing – original draft, Software, Methodology, Investigation, Funding acquisition. **Cristina López-Bravo:** Writing – original draft, Supervision, Investigation, Funding acquisition, Formal analysis. **Francisco Javier González-Castaño:** Writing – review & editing, Supervision, Funding acquisition, Conceptualization. **Felipe Gil-Castiñeira:** Writing – review & editing, Supervision, Resources, Funding acquisition. **David Candal-Ventureira:** Writing – review & editing, Visualization, Validation, Data curation.

**Declaration of competing interest**

The authors declare that they have no known competing financial interests or personal relationships that could have appeared to influence the work reported in this paper.

**Data availability**

We have open sourced the simulators developed and they are referenced and linked in the manuscript ([34] and [39])

**Acknowledgments**

This work was supported in part by Xunta de Galicia (Spain) under grants ED481B-2022-019 and ED431C 2022/04; grants PRE2021-098290, PID2020-116329GB-C21 and PDC2021-121335-C21 from the Ministry of Science and Innovation (MCIN), State Investigation Agency (AEI), Spain, and the European Social Fund (FSE+); and Universidade de Vigo/CISUG for open access.

**References**

[1] T.S. Rappaport, S. Sun, R. Mayzus, H. Zhao, Y. Azar, K. Wang, G.N. Wong, J.K. Schulz, M. Samimi, F. Gutierrez, Millimeter wave mobile communications for 5G cellular: It will work!, IEEE Access 1 (2013) 335–349.

[2] C. Han, Y. Wu, Z. Chen, X. Wang, Terahertz communications (TeraCom): Challenges and impact on 6G wireless systems, 2019, arXiv preprint arXiv:1912.06040.

[3] H. Elayan, O. Amin, B. Shihada, R.M. Shubair, M.-S. Alouini, Terahertz band: The last piece of RF spectrum puzzle for communication systems, IEEE Open J. Commun. Soc. 1 (2019) 1–32.

[4] C. Chaccour, M.N. Soorki, W. Saad, M. Bennis, P. Popovski, M. Debbah, Seven defining features of terahertz (THz) wireless systems: A fellowship of communication and sensing, IEEE Commun. Surv. Tutor. 24 (2) (2022) 967–993.

[5] C. Chaccour, M.N. Soorki, W. Saad, M. Bennis, P. Popovski, Can terahertz provide high-rate reliable low-latency communications for wireless VR? IEEE Internet Things J. 9 (12) (2022) 9712–9729.

[6] S. Mangiante, G. Klas, A. Navon, Z. GuanHua, J. Ran, M.D. Silva, VR is on the edge: How to deliver 360 videos in mobile networks, in: Proceedings of the Workshop on Virtual Reality and Augmented Reality Network, 2017, pp. 30–35.

[7] A. Shafie, N. Yang, S.A. Alvi, C. Han, S. Durrani, J.M. Jornet, Spectrum allocation with adaptive sub-band bandwidth for terahertz communication systems, IEEE Trans. Commun. 70 (2) (2021) 1407–1422.

[8] R. Liu, M. Lee, G. Yu, G.Y. Li, User association for millimeter-wave networks: A machine learning approach, IEEE Trans. Commun. 68 (7) (2020) 4162–4174.

[9] M.-T. Suer, C. Thein, H. Tchouankem, L. Wolf, Multi-connectivity as an enabler for reliable low latency communications—An overview, IEEE Commun. Surv. Tutor. 22 (1) (2019) 156–169.

[10] Y. Liu, X. Fang, M. Xiao, S. Mumtaz, Decentralized beam pair selection in multi-beam millimeter-wave networks, IEEE Trans. Commun. 66 (6) (2018) 2722–2737.

[11] I. Shayea, M. Ergen, A. Azizan, M. Ismail, Y.I. Daradkeh, Individualistic dynamic handover parameter self-optimization algorithm for 5G networks based on automatic weight function, IEEE Access 8 (2020) 214392–214412.

[12] 3GPP, TS36.331: Evolved universal terrestrial radio access (E-UTRA); radio resource control (RRC); protocol specification (release 9), 2016.

[13] G. Araniti, J. Cosmas, A. Iera, A. Molinaro, A. Orsino, P. Scopelliti, Energy efficient handover algorithm for green radio networks, in: 2014 IEEE International Symposium on Broadband Multimedia Systems and Broadcasting, 2014, pp. 1–6, http://dx.doi.org/10.1109/BMSB.2014.6873558.

[14] A.H. Arani, M.J. Omidi, A. Mehbodniya, F. Adachi, A handoff algorithm based on estimated load for dense Green 5G networks, in: 2015 IEEE Global Communications Conference, GLOBECOM, 2015, pp. 1–7, http://dx.doi.org/10.1109/GLOCOM.2015.7417634.

[15] L. Sun, J. Hou, T. Shu, Optimal handover policy for mmwave cellular networks: A multi-armed bandit approach, in: 2019 IEEE Global Communications Conference, GLOBECOM, 2019, pp. 1–6, http://dx.doi.org/10.1109/GLOBECOM38437.2019.9014079.

[16] 3GPP, TS37.340: NR; multi-connectivity; overall description; stage-2, 2017.

[17] A. Koutsaftis, M.F. Özkoç, F. Fund, P. Liu, S.S. Panwar, Fast wireless backhaul: A multi-connectivity enabled mmwave cellular system, in: GLOBECOM 2022 - 2022 IEEE Global Communications Conference, 2022, pp. 1813–1818, http://dx.doi.org/10.1109/GLOBECOM48099.2022.10001455.

[18] Y. Zhao, X. Zhang, X. Gao, K. Yang, Z. Xiong, Z. Han, Dual-connectivity handover scheme for a 5G-enabled ambulance, IEEE Trans. Commun. 71 (9) (2023) 5320–5334, http://dx.doi.org/10.1109/TCOMM.2023.3283770.

[19] S. Wu, M. Alrabeiah, C. Chakrabarti, A. Alkhateeb, Blockage prediction using wireless signatures: Deep learning enables real-world demonstration, IEEE Open J. Commun. Soc. 3 (2022) 776–796, http://dx.doi.org/10.1109/OJCOMS.2022.3162591.

[20] C. Lee, H. Cho, S. Song, J.-M. Chung, Prediction-based conditional handover for 5G mm-wave networks: A deep-learning approach, IEEE Veh. Technol. Mag. 15 (1) (2020) 54–62, http://dx.doi.org/10.1109/MVT.2019.2959065.

[21] M. Qurratulain Khan, A. Gaber, P. Schulz, G. Fettweis, Machine learning for millimeter wave and terahertz beam management: A survey and open challenges, IEEE Access 11 (2023) 11880–11902, http://dx.doi.org/10.1109/ACCESS.2023.3242582.

[22] Q. Xue, Y.-J. Liu, Y. Sun, J. Wang, L. Yan, G. Feng, S. Ma, Beam management in ultra-dense mmwave network via federated reinforcement learning: An intelligent and secure approach, IEEE Trans. Cogn. Commun. Netw. 9 (1) (2023) 185–197, http://dx.doi.org/10.1109/TCCN.2022.3215527.

[23] Y. Zhang, M. Alrabeiah, A. Alkhateeb, Reinforcement learning of beam codebooks in millimeter wave and terahertz MIMO systems, IEEE Trans. Commun. 70 (2) (2022) 904–919, http://dx.doi.org/10.1109/TCOMM.2021.3126856.

[24] A. Alkhateeb, I. Beltagy, S. Alex, Machine learning for reliable mmwave systems: Blockage prediction and proactive handoff, in: 2018 IEEE Global Conference on Signal and Information Processing, GlobalSIP, 2018, pp. 1055–1059, http://dx.doi.org/10.1109/GlobalSIP.2018.8646438.

[25] Y. Oguma, T. Nishio, K. Yamamoto, M. Morikura, Proactive handover based on human blockage prediction using RGB-D cameras for mmwave communications, IEICE Trans. Commun. 99 (8) (2016) 1734–1744.

[26] S. Wu, C. Chakrabarti, A. Alkhateeb, Proactively predicting dynamic 6G link blockages using LiDAR and in-band signatures, IEEE Open J. Commun. Soc. 4 (2023) 392–412, http://dx.doi.org/10.1109/OJCOMS.2023.3239434.

[27] G. Charan, M. Alrabeiah, A. Alkhateeb, Vision-aided 6G wireless communications: Blockage prediction and proactive handoff, IEEE Trans. Veh. Technol. 70 (10) (2021) 10193–10208, http://dx.doi.org/10.1109/TVT.2021.3104219.

[28] N. Abuzainab, M. Alrabeiah, A. Alkhateeb, Y.E. Sagduyu, Deep learning for THz drones with flying intelligent surfaces: Beam and handoff prediction, in: 2021 IEEE International Conference on Communications Workshops, ICC Workshops, 2021, pp. 1–6, http://dx.doi.org/10.1109/ICCWorkshops50388.2021.9473804.

[29] S. Dwivedi, R. Shreevastav, F. Munier, J. Nygren, I. Siomina, Y. Lyazidi, D. Shrestha, G. Lindmark, P. Ernström, E. Stare, et al., Positioning in 5G networks, IEEE Commun. Mag. 59 (11) (2021) 38–44.

[30] A. Shahmansoori, G.E. Garcia, G. Destino, G. Seco-Granados, H. Wymeersch, Position and orientation estimation through millimeter-wave MIMO in 5G systems, IEEE Trans. Wireless Commun. 17 (3) (2017) 1822–1835.

[31] N.A. Abbasi, A. Hariharan, A.M. Nair, A.F. Molisch, Channel measurements and path loss modeling for indoor THz communication, in: 2020 14th European Conference on Antennas and Propagation, Eucap, IEEE, 2020, pp. 1–5.

[32] K.M.S. Huq, J. Rodriguez, I.E. Otung, 3D network modeling for THz-enabled ultra-fast dense networks: A 6G perspective, IEEE Commun. Stand. Mag. 5 (2) (2021) 84–90.

[33] Z. Hossain, Q. Xia, J.M. Jornet, TeraSim: An ns-3 extension to simulate terahertz-band communication networks, Nano Commun. Netw. 17 (2018) 36–44.

[34] P. Fondo-Ferreiro, C. López-Bravo, F.J. González-Castaño, F. Gil-Castiñeira, D. Candal-Ventureira, BloTHz: THz network blockage simulator, 2024, http://dx.doi.org/10.5281/zenodo.10681711.

[35] R. Singh, D. Sicker, An analytical model for efficient indoor THz access point deployment, in: 2020 IEEE Wireless Communications and Networking Conference, WCNC, IEEE, 2020, pp. 1–8.

[36] 3GPP, Study on Channel Model for Frequency Spectrum Above 6 GHz (3GPP TR 38.900 Version 15.0.0 Release 15), Technical Report (TR) 38.900, 3rd Generation Partnership Project (3GPP), 2018, Version 15.0.0.

[37] M. Anas, F.D. Calabrese, P.E. Mogensen, C. Rosa, K.I. Pedersen, Performance evaluation of received signal strength based hard handover for UTRAN LTE, in: 2007 IEEE 65th Vehicular Technology Conference, VTC2007-Spring, IEEE, 2007, pp. 1046–1050.

[38] K. Sarabandi, A. Jam, M. Vahidpour, J. East, A novel frequency beam-steering antenna array for submillimeter-wave applications, IEEE Trans. Terahertz Sci. Technol. 8 (6) (2018) 654–665.

[39] P. Fondo-Ferreiro, THz visibility prediction and AP assignment simulator, URL https://gti-uvigo.github.io/THz-visibility-prediction-and-AP-assignment-simulator. (Accessed 14 May 2024).